# On slowdown of particles in a vacuum at their delocalization from an electromagnetic well with a depth decreasing with time

## A.Ch. Izmailov

*Institute of Physics, Azerbaijan National Academy of Sciences, Baku, AZ-1143, Azerbaijan*

*e-mail:* azizm57@rambler.ru

## Abstract

The possibility of a significant slowdown of particles by removing them from a localized state in an electromagnetic potential well with a fixed spatial distribution is shown with a sufficiently slow decrease in the depth of this well with time. It is believed that the considered particles are in high vacuum conditions and forces acting on them are not dissipative, i.e. these particles move without friction. Depending on whether the particles have an electric (magnetic) dipole moment, a controlled electric (magnetic) field or nonresonant laser radiation can be used to slow them down in this way. A detailed analysis of the features and efficiency of the established particle deceleration mechanism was carried out on the example of the one-dimensional rectangular potential well based on simple mathematical relationships. The obtained results can be used in the mechanics of micro- and nanoparticles in high vacuum, and under certain conditions also in ultra-high resolution spectroscopy of atoms and molecules "cooled" by the proposed method.

## 1. INTRODUCTION

The development of effective methods for slowdown (cooling) of various micro- and nanoparticles in vacuum is important for studying the fundamental problems of gravity and quantum mechanics using these particles, including limits of applicability of the corresponding theories [1, 2]. In addition, such particles can be the basis for creating more accurate accelerometers and magnetometers, as well as new promising composite materials. At the same time, deep cooling of atoms and molecules in vacuum is necessary for precision fundamental research [3].

The author's articles [4, 5] showed the possibility of slowdown and trapping of various particles (including atoms and molecules in their ground quantum state) by means of external electromagnetic fields, which induce potential wells on the path of such particles with a fixed spatial distribution, but deepening with time up to some limit. It is assumed that the considered particles are in high vacuum conditions and forces acting on them are not dissipative, i.e. these particles move without friction. Depending on whether the particles have an electric (magnetic) dipole moment, a controlled electric (magnetic) field can be used to capture or slow them down by the proposed methods [4, 5]. Of particular note is the possibility of creating the considered potential wells of various fixed spatial configurations from a single nonresonant laser beam with controlled intensity by converting this beam by various stationary optical elements (mirrors, lenses, polarizers, prisms). In this case, a situation should be realized when the light pressure force from the side of the given radiation is negligibly small compared to the light-induced gradient force acting on the particle under study [6].

This brief communication shows the possibility of significant slowdown (cooling) of such particles in a vacuum due to their removal from the initial localized state in a similar potential well with a sufficiently slow decrease in the depth of this well by means of a controlled weakening of the corresponding electromagnetic field. The following section 2 presents the general relationships and reasoning for the study of the considered processes. Further, in Section 3, based on very simple mathematical formulas, the detailed analysis of the possible significant slowdown (cooling) of particles compared to their initial state is carried out using the example of the one-dimensional rectangular potential well. In conclusion, the findings of the work are presented.

## 2. BASIC RELATIONSHIPS

As in articles [4,5], we will conduct theoretical study within the framework of classical mechanics and electrodynamics. Let us assume that a point particle with mass $m$ moves in vacuum, including the region of the potential well $U(\mathbf{r}, t)$, which explicitly depends not only on the coordinate $\mathbf{r}$, but also on the time $t$. The total energy of the particle with a nonrelativistic velocity $v$ is described by the well-known formula [7]:

$$E(\mathbf{r}, \mathbf{v}, t) = 0.5mv^2 + U(\mathbf{r}, t). \qquad (1)$$

Next, we will consider the following potential energy $U(\mathbf{r}, t)$:

$$U(\mathbf{r}, t) = s(\mathbf{r}) * \varphi(t), \qquad (2)$$

where the coordinate function $s(\mathbf{r}) \leq 0$ and $\varphi(t)$ is a function of time $t$. Such a non-stationary potential (2) can be created for particles with an electric or magnetic moment by means of a controlled electromagnetic field (in particular, non-resonant laser radiation) with a time-varying intensity, but with a fixed spatial distribution [4, 5]. In this case (2), we obtain the following equation of particle motion [7]:

$$m \frac{d^2 \mathbf{r}}{dt^2} = -\varphi(t) \frac{ds(\mathbf{r})}{d\mathbf{r}}. \qquad (3)$$

We consider the situation when the force on the right side of equation (3) does not have a dissipative effect on the motion of the particle, i.e. is not a force of friction. Then relations (1)-(3) directly imply an important formula for the time derivative of the total particle energy $E$:

$$\frac{dE}{dt} = s(\mathbf{r}) \frac{\varphi(t)}{dt}. \qquad (4)$$

In articles [4, 5], situations were considered when particles fall into a potential well (2), which deepens with time. Then, according to (4), the total energy of the particle $E$ (1) decreases, since the function $s(\mathbf{r}) \leq 0$, and the derivative $\frac{\varphi(t)}{dt} \geq 0$. Therefore, particles with relatively low initial velocities can be trapped in such a well that deepens with time, and faster particles overcome this well, losing part of their initial kinetic energy, i.e. are slowing down.

In this paper, we will consider the situation when a particle with the total energy $E_0 < 0$ is preliminarily localized in an electromagnetic potential well, which is stationary until a certain time $t = t_0$. However, further at $t > t_0$, the depth of this well (2) decreases sufficiently slowly by means of a controlled weakening of the corresponding external electromagnetic field. Then in Eq.(4) the derivative $\frac{\varphi(t)}{dt} < 0$ and by some time $t = t_1$ the total energy (1) of the localized particle will rise to zero $E(t_1) = 0$. Subsequently, having reached the nearest boundary of the potential well (2) at the moment $t = t_2$, this particle can already go beyond it, with a relatively low final velocity $v_f$ and total energy $E_f$. This process is clearly shown in Fig. 1 using the example of a one-dimensional well. In this way, with a sufficiently slow decrease in the depth of the potential well (2), it is possible to achieve a significant "cooling" of the particles compared to their initial localized state.

The preliminary stage of particle localization, which is necessary for the process under consideration, can be realized as a result of the capture of this particle by the same well (2) with its sufficiently rapid deepening with time up to a certain limit [4, 5]. In the general case, to conduct an appropriate theoretical study, it is necessary to numerically solve the differential equations of particle motion (3). However, a number of important characteristic features and the efficiency of slowdown (cooling) of particles by the proposed method will be further established using the example of a one-dimensional rectangular well based on simple mathematical relationships.

## 3. RECTANGULAR WELL

Consider the rectangular potential well $U(x,t)$, bounded by the coordinates $x = \pm L$, whose depth decreases with time $t$ (Fig. 2):

$$U(x,t) = -J_0 \eta(L^2 - x^2)\varphi(t), \qquad (5)$$

where $J_0 > 0$ is the constant with the dimension of energy, $\eta(L^2 - x^2)$ is the step function ($\eta(y) = 1$ for $y \geq 0$ and $\eta(y) = 0$ for $y < 0$), $0 \leq \varphi(t) \leq 1$ is a non-increasing function of time. It is believed that until the moment $t = t_0$ the potential well (5) is stationary with the value $\varphi(t) = 1$ and a particle with mass $m$ and velocity $v_0$ is localized in it. Then, taking into account (5), the total energy of this particle at $t \leq t_0$ has the form:

$$E_0 = 0.5 m v_0^2 - J_0. \qquad (6)$$

Localization of a particle in a well (5) is possible only at the energy $E_0 \leq 0$ (6), i.e. under the following limitation on the particle velocity $v_0$:

$$|v_0| \leq v_m = \sqrt{\frac{2J_0}{m}}. \qquad (7)$$

Starting from the moment $t = t_0$, the depth of the well (5) starts to decrease. Such a process is accompanied by an increase in the total particle energy $E(t)$, according to relation (4), which in the case of a rectangular well (5) reduces to the following formula for $t \geq t_0$:

$$E(t) - E_0 = J_0[\varphi(t_0) - \varphi(t)] = J_0[1 - \varphi(t)]. \qquad (8)$$

From (5)-(8) we obtain the condition under which the particle energy $E(t_1) = 0$ at some time $t_1 > t_0$:

$$\varphi(t_1) = \frac{m v_0^2}{2 J_0} = \frac{v_0^2}{v_m^2} \leq 1. \qquad (9)$$

According to the equation of motion (3), the module $|v_0|$ (7) of the particle velocity will be constant within the one-dimensional rectangular potential well (5) despite the change in its depth. At $t \geq t_1$ this particle can already leave this well after a time interval $\Delta t = \Delta x/|v_0|$, where $0 \leq \Delta x \leq 2L$ is the distance traveled from the point $x(t_1)$ towards the corresponding boundary of the well with coordinates $L$ or $-L$ (Fig. 2). Coordinate $x(t_1)$ depends on position $x(t_0)$ of the considered particle at the initial time $t = t_0$. The final energy $E_f$ and particle speed $v_f$ at the exit from the well (5) is determined on the basis of relation (4) taking into account the fact that at the moment $t = t_1$ total energy $E(t_1) = 0$:

$$E_f = E\left(t_1 + \frac{\Delta x}{|v_0|}\right) - E(t_1) = J_0\left[\varphi(t_1) - \varphi\left(t_1 + \frac{\Delta x}{|v_0|}\right)\right] = 0.5mv_f^2. \quad (10)$$

From (9) and (10) we obtain the following ratio $r$ of the particle final velocity modulus $|v_f|$ to the modulus of its speed $|v_0|$ in the previous time of its localization in this well:

$$r = \frac{|v_f|}{|v_0|} = \sqrt{1 - \frac{\varphi\left(t_1 + \frac{\Delta x}{|v_0|}\right)}{\varphi(t_1)}}. \quad (11)$$

The value $\varphi(t_1)$ and time $t_1$ in (11) are determined from relation (9). Ratio $r$ (11) grows with increasing distance $\Delta x$ from 0 to $2L$ due to the decrease in time of the function $\varphi(t)$ (5), and the speed $v_f = 0$ if $\Delta x = 0$. Indeed, at $\Delta x = 0$ the total energy of the particle $E(t_1) = 0$ occurs at the point $x(t_1)$, which coincides with one of the boundaries of the well $x = \pm L$ (Fig. 2) where the exit of the particle is carried out at the moment $t_1$.

Based on formula (11), we also introduce the following characteristic $\sigma$, which determines the distance-averaged $(0 \leq \Delta x \leq 2L)$ the ratio of the values of the kinetic energies of a particle when it leaves the potential well and in the process of its previous localization in this well:

$$\sigma = \langle r^2 \rangle = \frac{1}{2L}\int_0^{2L}\left[1 - \frac{\varphi\left(t_1 + \frac{\theta}{|v_0|}\right)}{\varphi(t_1)}\right]d\theta. \quad (12)$$

Expressions for quantities $r$ (11) and $\sigma$ (12) are significantly simplified for the exponential function $\varphi(t)$ in (5), which can be represented as:

$$\varphi(t) = \exp\left[-\left(\frac{t-t_0}{T}\right)\right]\eta(t - t_0) + \eta(t_0 - t), \tag{13}$$

where $T$ is some characteristic time interval, $\eta(y)$ is the step function. After substitution of the function $\varphi(t)$ (13) in relations (11) and (12) we obtain the following simple formulas:

$$r = \frac{|v_f|}{|v_0|} = \sqrt{1 - \exp\left(-\frac{\Delta x}{|v_0|T}\right)}, \tag{14}$$

$$\sigma = 1 - \frac{|v_0|T}{2L}\left[1 - \exp\left(-\frac{2L}{|v_0|T}\right)\right]. \tag{15}$$

Fig. 3 shows the dependencies of the value $\langle r \rangle$ obtained from the expression $r$ (14) after its averaging over the distance $0 \leq \Delta x \leq 2L$, as well as the value of $\sigma$ (15) on the speed $v_0$ of a particle previously localized in a rectangular well (5). These quantities monotonically decrease with increasing $v_0$ starting from 1 at $v_0 = 0$ to their minimum values at $v_0 = v_m$ (7). The velocity $v_0$ on the abscissa axis in Fig. 3 is presented in units of $w = 2L/T$, where $2L$ is the length of the potential well (Fig. 2), and $T$ is the characteristic time interval for the function $\varphi(t)$ (13) at $t \geq t_0$. According to formulas (14), (15) and Fig. 3, a multiple decrease in the kinetic energy of a particle compared to its value in the initial localized state is possible under the following condition:

$$\frac{2L}{T} \ll |v_0| \leq \sqrt{\frac{2J_0}{m}}. \tag{16}$$

Thus, the efficiency of the proposed method of slowing down (cooling) of particles increases with a longer decrease in the depth of the potential well, as well as with a reduction in its length. It is important to note that, according to condition (16), for the characteristic time $T$ function decreasing $\varphi(t)$ (5), (13) quite a lot of oscillations of the particle between the boundaries of the well should occur.

## 4. CONCLUSION

In this paper, we show the possibility of a significant slowdown (cooling) of particles in a high vacuum by means of their delocalization from a potential well (2) created by an external electromagnetic field with a fixed spatial distribution. To do this, two controlled processes of changing the depth of a well of this type (2) with time are sequentially implemented. First, to capture a passing particle, it is necessary to carry out a sufficiently rapid deepening of this well on its way by amplifying the external electromagnetic field to a certain limit [4, 5]. Further, the process of delocalization of this particle considered by us occurs with a relatively slow decrease in the depth of this well. A detailed study of such a particle deceleration process was carried out on the example of the one-dimensional rectangular well based on simple mathematical relations. In practice, such a well corresponds, in particular, to a region of a local quasi-homogeneous electric (or magnetic) field created along the direction of particle flight with a dipole electric (or magnetic) moment.

At the same time, numerical calculations performed by the author based on the equations of motion of particles (from Section 2) showed that the qualitative results obtained in this work are also valid for one-dimensional potential wells of limited length with other shapes (in particular, Gaussian). Thus, it has been established that for a significant slowdown (cooling) of particles by the proposed method, such a long process of reducing the depth of such an electromagnetic well is necessary, during which quite a lot of oscillations of pre-localized particles between the boundaries of this well should occur (Fig. 1). Based on the simple formula for the value $\sigma$ (15), one can estimate the effectiveness of this method by substituting there instead of $2L$ the characteristic value of the length of the corresponding one-dimensional well, and instead of $|v_0|$ the average modulus of the velocity of a localized particle over the period of its oscillations before the beginning of the decrease in the depth of this well according to the dependence $\varphi(t)$ (13). Note that the one-dimensional potential wells necessary for such processes can also be created by light radiation. In particular, such a well occurs for particles moving in the orthogonal direction through the center of a nonresonant Gaussian beam [3, 6] with controlled intensity.

It should be noted that in order to analyze the features of such a slowdown of particles in the case of two- and three-dimensional potential wells of type (2), it is necessary to carry out additional theoretical studies.

Despite the fact that the present work was carried out within the framework of classical mechanics and electrodynamics, its results, under certain conditions, are also applicable to atoms and molecules. Indeed, potential wells of type (2) can be created for these particles in their ground quantum state by means of an intensity-controlled laser radiation with a fixed spatial distribution and with a frequency substantially detuned from resonances with atomic (molecular) transitions [3, 6].

Thus, the results of this work can be used not only in the mechanics of micro- and nanoparticles in high vacuum, but also in ultrahigh-resolution spectroscopy of atoms and molecules "cooled" by the proposed method.

______________________________________________

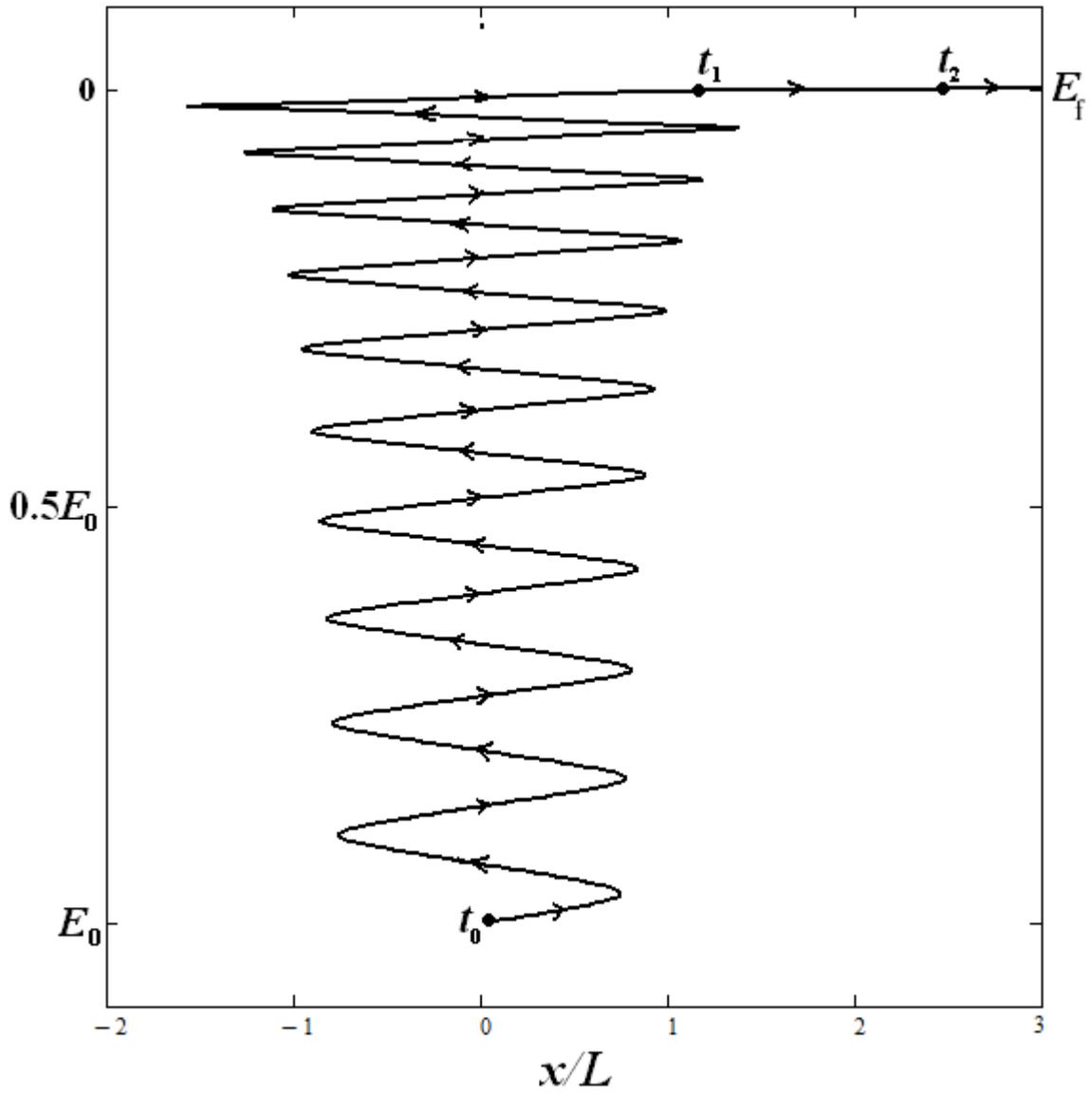

**Fig.1.** Scheme of changing the $x$ coordinate and total energy $E$ of a particle during its delocalization from a one-dimensional Gaussian-shaped well (with a characteristic length of $2L$) by means of a sufficiently slow decrease in the depth of this well with time $t$, starting from the particle energy $E_0 < 0$ at the moment $t = t_0$. Then, at the time $t_1 > t_0$, the energy of the particle $E(t_1) = 0$, and finally, at the time $t_2 > t_1$, the particle with the final energy $E_\mathrm{f} \geq 0$ goes beyond the influence of this well.

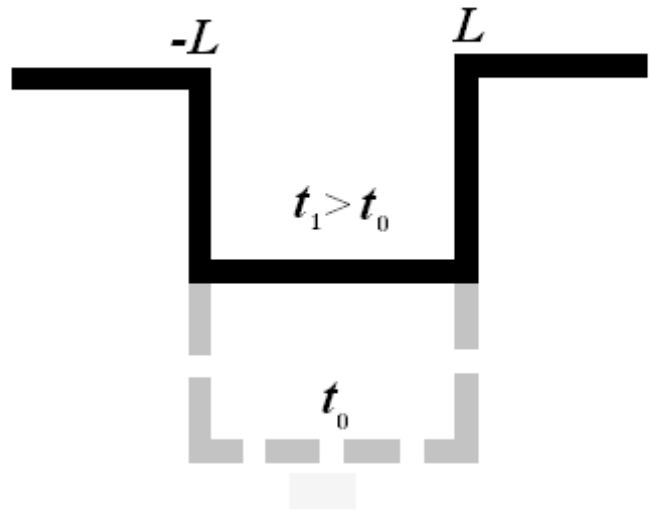

**Fig.2.** One-dimensional rectangular potential well with depth decreasing with time $t$.

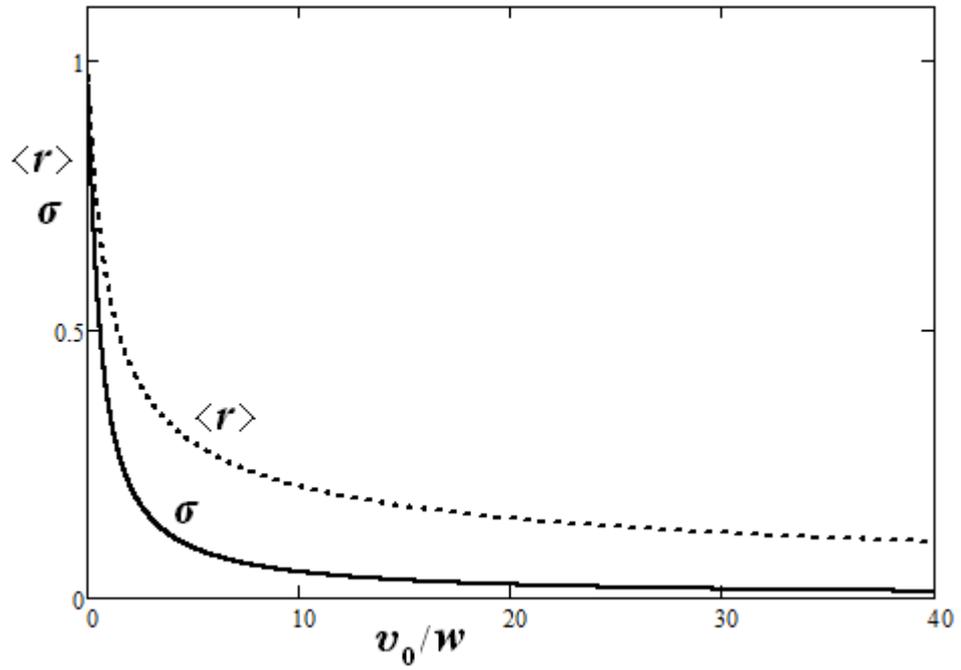

**Fig.3.** The dependence of the quantities $\langle r \rangle$ and $\sigma$ on the particle velocity $v_0$ (in units of $w = 2L/T$) in the case of rectangular potential well for the function $\varphi(t)$ (13) at $t \geq t_0$, when $v_m = \sqrt{2J_0/m}$=40w.